\documentclass[sigchi]{acmart}

\usepackage{booktabs} 
\usepackage{enumitem}

\acmConference[Virginia Tech Workshop on the Future of Human-Computer Interaction]{Virginia Tech Workshop on the Future of Human-Computer Interaction}{April 2019}{Tucson, Virginia, USA}
\acmYear{2019}
\copyrightyear{2019}

\begin{document}
\title{Computer-mediated Empathy}

\author{Sang Won Lee}
\orcid{1234-5678-9012}
\affiliation{%
  \institution{Computer Science}
  \institution{Virginia Tech}
}
\email{sangwonlee@vt.edu}

\renewcommand{\shortauthors}{S. Lee}
\newcommand{\out}[1]{{#1}}
\newcommand{\sang}[1]{\out{{\small\textcolor{red}{\bf [Sang: #1]}}}}

%
%

\maketitle

\section{Introduction}
Today, we live in an era in which we can communicate via computers more than ever before. 
While novel social networks and emerging technologies help us transcend the spatial and temporal constraints inherent to in-person communication, the trade-off is a loss of natural expressivity. While empathetic interaction is already challenging in in-person communication, computer-mediated communication makes such empathetically rich communication even more difficult. Are technology and intelligent systems opportunities or threats to more empathic interpersonal communication? 

My future research vision is to build \textit{computational systems that facilitate understanding and empathy}. 
Realizing empathy is suggested not only as a way to communicate with others, but also to design products for users and facilitate creativity~\cite{lim2013realizing}. In this position paper, I suggest a framework to breakdown empathy, introduce each element, and show how computing, technologies, and algorithms can support (or hinder) certain elements of the empathy framework. 

\section{Empathy Framework}
Empathic interactions involve two roles --- the empathizer and the empathizee. Figure 1 shows the framework of empathy. 
\begin{figure}
    \centering
    \includegraphics[width=\columnwidth]{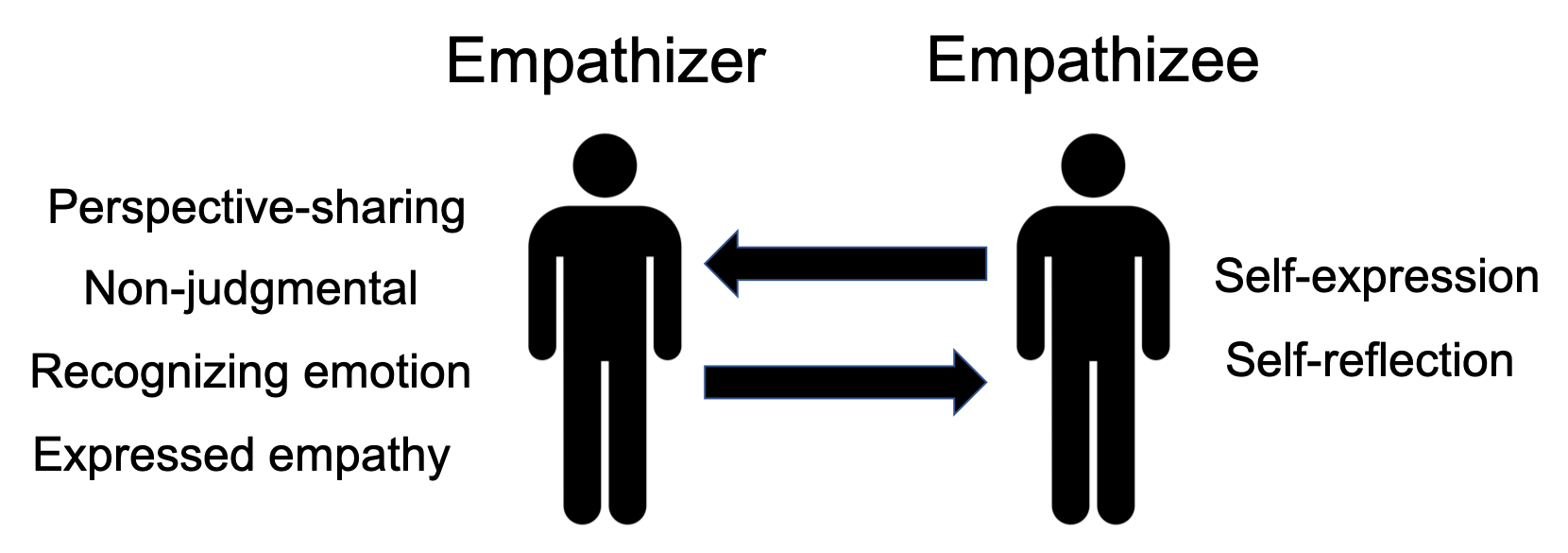}
    \caption{Empathy framework: Empathy is a dyadic interaction between empathizer and empathizee, each requiring a different set of supports}
    \label{fig:my_label}
\end{figure}
\subsection{Empathizee}
Empathy is often reduced to a problem in terms of one person needing to empathize with another. 
However, the other  --- the person who eventually receives empathy --- is a significant stakeholder in this relation, and is often called the empathizee~\cite{gladstein1987all}. 
Some anthropologists view empathy as depending on what empathizees  are ``willing or able to tell about themselves.''~\cite{}. 
Seeing the dynamic as dyadic emphasizes the importance of the person to be empathized with, placing that person on par with the person displaying empathy.

\textbf{(Self-expression)} How can we help empathizees with computing? 
The empathizee does not need to be present to the empathizers and can instead be ``deeply heard from sound or video recording or perhaps even from written expression, an artistic work, or another expressive product~\cite{barrett1981empathy}.'' 
This highlights the need for the empathizee to be afforded an expressive medium, including and beyond in-person communication --- any kind of creative (or even artistic) practice through which one can express their thoughts and emotions may suffice. 
This can include simply computational supports in creative practice; namely writing, art, music, research, and other novel expressive media. 
Computational tools for self-expression can help constitute an empathic attention set~\cite{barrett1981empathy}.

\textbf{(Self-reflection)} Self-reflection is a process that may need to be followed by self-expression. 
It is the exercise of introspection, coupled with the willingness to learn about oneself, in order to help achieve self-awareness. 
Therefore, self-reflection is important to understanding what needs to be expressed. 
Furthermore, self-reflection is something that can support empathizers, as an understanding of themselves leads to enhancements in the ability to empathize with others~\cite{knowthyselves}. 
Therefore, computational systems that promote self-reflection can enhance people's abilities to empathize with others. 

\subsection{Empathizer}
Psychologists typically consider empathy as an individual ability --- the empathizer's ability ---  to share others' feelings by observing or learning about their emotional state~\cite{decety2006human}.
Wiseman finds that there are four defining attributes of empathy: 1) perspective taking (``see the world as others see it''), 2) non-judgmental, 3) understanding another's feelings, and 4) communicating that understanding~\cite{wiseman1996concept}. 

\textbf{(Perspective Sharing)} Being able to take another's perspective --- a cognitive function --- is an essential part of empathy. 
As people often express empathy as ``being in someone else's shoes'', the process of realizing empathy emphasizes sharing the visual perspective of the other. 
Therefore, \textbf{perspective taking} is the one of the essential elements in empathizing with others. 
I believe emerging mixed reality technologies can be used to support perspective taking, as with augmented reality and virtual reality --- one can literally render the other's perspective. 
For example, virtual reality can put a empathizer in an empathizee's specific situations (e.g., homelessness) ~\cite{herrera2018building}. 
Researchers have used AR/VR technologies to facilitate empathic interaction between remote collaborators~\cite{lee2016remote,lee2017improving}. 
Lastly, one needs to distinguish empathic response from own reactions drawing on prior experience and triggered by the empathizee's perspectives.
This includes recognizing other qualities that may have affected how the empathizee works, which may include the empathizee's environment, personal traits, and other relevant contextual information. 
Alternatively, it may include unfolding information over the temporal dimension that empathizers may have not access to in real time~\cite{sang_won_lee_2018_1471026}. 
To that end, the empathizee will benefit from context-aware computing environments that can supplement expressiveness in regard to their experience~\cite{4624429}.

\textbf{(Non-judgmental)} Being non-judgmental is an area in which computing can help empathizers, as computers is not judgmental by default without any algorithm injected by human. Human-to-human communication can be mediated in a way that is less judgmental. For example, asynchronous textual communication enabled by computers can promote empathic communication by blocking visual cues that could have prompted judgment in in-person face-to-face conversation. Communication may also be moderated by algorithms to keep the conversation neutral before transmission. 

\textbf{(Recognizing Emotions)} Emotion recognition is a long-standing topic in the field of affective computing and natural language processing. For example, speech and facial expressions have been analyzed for computers to understand the emotions of others. Therefore, existing research in these fields can support empathetic conversation, particularly when these modalities available in in-person communication are not available in other types of computer-mediated communication. Thus, enriching the communication medium may be a means of restoring or augmenting emotional recognition in computational media. Other types of communication can be similarly augmented with emotion recognition techniques. 

\textbf{(Communicating Understanding)} Empathizers communicating back to empathizees to express empathy is a vital element of the empathy cycle~\cite{barrett1981empathy}. 
In computer-mediated empathy, as empathic interaction can happen without empathizees realizing it, awareness of empathy can help, closing the loop of empathy. 
In addition, empathizees being aware of the received empathy is often followed by further, deeper conversation on a subject that can further reinforce empathy.

\section{Computational Empathy and Computational Thinking}

In the suggested framework, empathizers and empathizees do not necessarily need to be human. Either end can be replaced with computers, machines, or algorithms. The case of the empathizer being a computer opens up a new field of \textbf{computational empathy} in which we make computers empathize with humans~\cite{paiva2017empathy}. This is largely is the most relevant scenario in the existing field of affective computing. However, additional arguments made for both empathizers and empathizees remain valid. For example, as a human empathizee, we need to express ourselves in ways that computers can easily understand. 

By contrast, computers can be empathizees, where a human must understand a computer's intentions. This is relevant to \textbf{computational thinking,} an essential skill for the immediate future, in which humans collaborate and interact with computers ubiquitously~\cite{wing2006computational}. Again, as with computers as empathizees, it is important to be able to express computer algorithms in human-comprehensible ways, which is relevant to the recent initiative of explainable artificial intelligence~\cite{gunning2017explainable}.

\bibliographystyle{ACM-Reference-Format}
\bibliography{sample-bibliography}

\end{document}